\documentclass[floatfix,aps,unsortedaddress,superscriptaddress,showpacs]
{revtex4}
\usepackage[dvips]{graphicx}
\usepackage{epsfig}

\begin{document}
\title{Completeness of the Coulomb scattering wave functions.}

\author{A.\,M.~Mukhamedzhanov}
\affiliation{Cyclotron Institute, Texas A\&M University,
College Station, TX 77843}

\author{M. Akin}
\affiliation{University of Georgia, Athens, GA}

\date{\today}
                               
\begin{abstract}
Completeness of the eigenfunctions of a self-adjoint Hamiltonian, which is the basic ingredient of quantum mechanics, plays an important role in nuclear reaction and nuclear structure theory. However, until now, there was no a formal proof of the completeness of the eigenfunctions of the two-body Hamiltonian with the Coulomb interaction. Here we present the first formal proof of the completeness of the two-body Coulomb scattering wave functions for repulsive unscreened Coulomb potential. To prove the completeness we use the Newton's method [R. Newton, J. Math Phys., 1, 319 (1960)]. The proof allows us to claim that the eigenfunctions of the two-body Hamiltonian with the potential given by the sum of the repulsive Coulomb plus short-range (nuclear) potentials also form a complete set. It also allows one to extend the Berggren's approach of modification of the complete set of the eigenfunctions by including the resonances for charged particles. We also demonstrate that the resonant Gamow functions with the Coulomb tail can be regularized using Zel'dovich's regularization method.
\end{abstract}

\maketitle  

\section{Introduction}
\label{section_1}
\begin{figure}[tbp]
\includegraphics*[width=\linewidth]{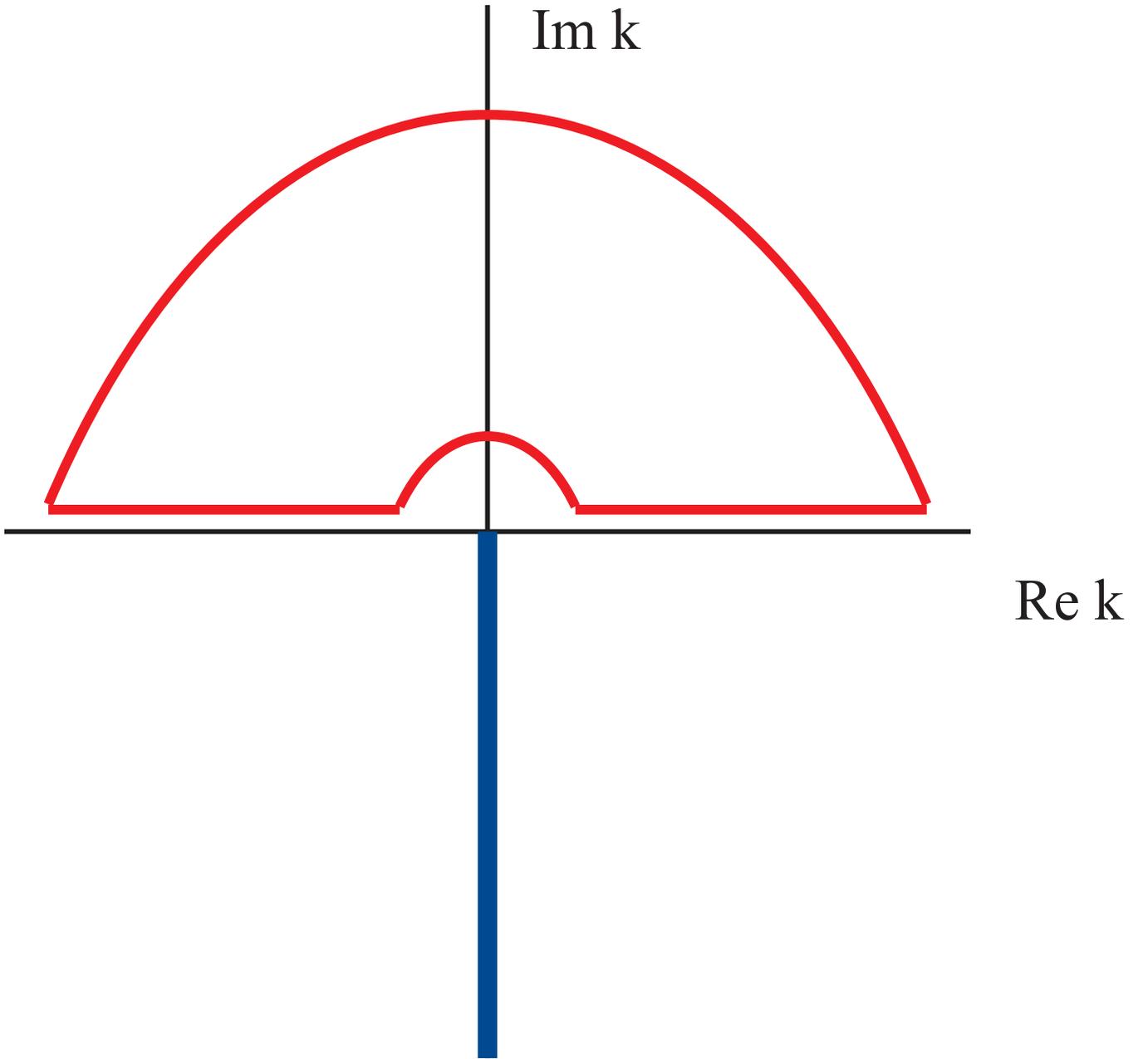}
\caption{(Color online). The red solid line is the Newton's integration contour over the closed contour $R$ in the complex $k$ plane. The blue solid line is the cut for the singular solution $f^{(+)}_{l}(k,r)$ .}  
\label{fig_1}
\end{figure}

Completeness of the eigenfunctions of a self-adjoint Hamiltonian as a part of the linear superposition principle provides a powerful tool in different areas of quantum physics. 
Consider the set of the radial eigenfunctions 
${\it S}=\{\varphi_{nl}(r),\,\psi_{l}(k,r)\},\;0 <n < N,\, 0 < k < \infty$ of the  of the self-adjoined Hamiltonian 
$H= T +V_{l}$. Here, $T$ is the kinetic energy  operator, $V_{l}(r)= V^{N}(r) + V_{l}(r)$ is the sum of the short-range and centrifugal potentilas,  
$\varphi_{nl}(r)$ are the normalized to unity eigenfunctions of the discrete branch of the energy spectrum (bound states), $n$ is the principal quantum number and $l$ is the orbital angular momentum; $\psi_{l}(k,r)$ are the eigenfunctions of the continious spectrum (scattering states) normalized to delta functions. 
For simplicity interacting particles are assumed to be spinless.
Definition of the completeness: set $\it S$ of the eigenfunctions  of the self-adjoined Hamiltonian forms a complete set in the Hilbert space \footnote[1]{For the continuum spectrum eigenfunctions are not square-integrable strictly speaking we need to use a rigged Hilbert space which extends the normal Hilbert space by bringing together the discrete and continuum spectrum eigenstates} if     
any function $h(r)$ belonging to this Hilbert space can be expanded in terms of the eigenfenctions: 
\begin{eqnarray}
 h(r) = \sum\limits_{n = 0}^N {C_{nl} \varphi _{nl} (r)}  + \int\limits_0^\infty  {dk\,} C_{l}(k)\psi _{kl} (r) \\ 
  = \int\limits_0^\infty  {dr'h(r')\,[\sum\limits_{n = 0}^N {\varphi _n^* (r')} \varphi _n (r)}  + \int\limits_0^\infty {\rm d}k\psi^{*}_{kl} (r')\, \psi_{l}(k,r)], 
\label{hfunction1}  
\end{eqnarray}
with
\begin{equation}
C_{nl} = \int\limits_0^\infty\,{dr\,r^{2}\,\varphi _{nl}(r)}\, h(r),
\label{cn1}
\end{equation}
\begin{equation}
C_{l}(k) = \int\limits_0^\infty\,{dr\,r^{2}\,\psi_{kl}^{*}(r)}\, h(r)
\label{clk1}
\end{equation}
and the norm
\begin{equation}
N= \int\limits_0^\infty\,\,{\rm d}r\,r^{2}\,h^{*}(r)\,h(r) < \infty.
\label{norm1}
\end{equation}
An elegant proof of the completeness of the eigenfunctions of the two-body Hamiltonian with the interaction potential $V^{N}(r)$ satisfying the conditions 
\begin{equation}
\int\limits_0^\infty  {dr\,r|V^{N}(r)| < \infty } 
\label{potcond1}
\end{equation}
and 
\begin{equation}
\int\limits_0^\infty  {dr\,r^{2l + 2} |V^{N}(r)| < \infty } 
\label{potcond2}
\end{equation}
has been presented long ago by R. Newton \cite{newton60,newton}. 
The first condition guarantees that a regular solution is entire function of $k$ and the second one secures behavior of the eigenfunctions at $k=0$.
Newton's method is based on consideration of the integral taken along the closed contour:      
\begin{equation}
I(r) = \oint\limits_{(C)}\,{\rm d}k\,k\,\int\limits_0^\infty\,{\rm d}r'\,h(r')\,G_l^{( + )} (r,r';k).  
\label{newtintegr1}
\end{equation}
Here, $G_l^{( + )} (r,r';k)$ is the two-body partial wave Green's function in the coordinate space. The closed integration contour $C$ in Eq. (\ref{newtintegr1}) is shown in Fig. \ref{fig_1}. It goes along the real axis in the momentum plane $k$ bypassing the origin $k=0$ along the infinitesimal contour $\gamma$, and along the semicircle $R$ with the radius $|k|_{R} \to \infty$. 

Although nobody doubts that the completeness holds also for the eigenfunctions of the Hamiltonian containing the unscreened Coulomb potential, a formal proof of it has not yet been presented in the literature. The reason for that is a quite complicated analytical behavior of the Coulomb Jost solutions $f^{(\pm)}_{l}(k,r)$ \footnote[2]{Jost solutions of the Schr\"odinger equation are singular at $r=0$ solutions in the complex $k$ plane asymptotically behaving as outgoing (for $(+)$ or ingoing (for $(-)$) waves.}. For the Coulomb interaction $f^{(+)}_{l}(k,r)$ ($f^{(-)}_{l}(k,r)$) has singularity at $k=0$ with the cut along the negative (positive) imaginary semiaxis in the $k$ plane \cite{ment65}. It makes behavior of the Jost functions and Jost solutions in a vicinity of $k=0$ quite complicated. Note that introducing the screened Coulomb potential
or a box changes the analytical properties of the Jost solutions: the singularity at $k=0$ dissapears (if the potential does not support a bound state at $E=0$) and the Coulomb cut along the imaginary semiaxis begins from $k= -i\,\frac{1}{R}$ ($k= i\,\frac{1}{R}$) for $f^{(+)}_{kl}(r)\;(f^{(-)}_{kl}(r))$, where $R$ is the screening radius, as it takes place for short range potentials. The Coulomb analytical features can be recovered only at $R=\infty$. 
   
In this paper we present, for the first time, a formal proof of the 
completeness of the eigenfunctions of the radial two-body Hamiltonian with pure repulsive Coulomb potential at arbitrary angular orbital momentum.
Our proof is based on the consideration of the Newton integral \cite{newton60}. The idea of the Newton's proof, should be applied to the pure Coulomb case, is as folloows: according to the Cauchy's  theorem the integral over a closed contour $C$, Eq. (\ref{newtintegr1}), which does not contain any singularities of the integrand on or inside the contour is zero. Then we need to show that the integral over large $R$ in he limit $|k|_{R} \to \infty$ gives $\pi\, h(r)$. This result is similar to the short range case. The most delicate problem is to prove that the integral over small semicircle $\gamma$ disappears in the limit $|k|_{\gamma} \to 0$, where $|k|_{\gamma}$ is the radius of the contour $\gamma$ in the momentum plane. The last step needed to prove the completeness is to show that the integral over the real axis can be reduced to 
$- \pi \int\limits_0^\infty dr'h(r')\, \int\limits_0^\infty  {\rm d}k\,\psi _{l}^{(C)*}(k,r')\,\psi_{l}^{(C)}(k,r)$. 

\section{General equations} 

From Fig. \ref{fig_1} it is clear that the closed contour $C$ in the complex $k$ plane consists of the different parts: $C=R+0+\gamma$. Here $R$ is the integration contour over the large semicircle,
$0$ is the integration contour along the real axis with eliminated small interval around $k=0$ and $\gamma$ is the integration contour over the small semicircle around $k=0$. 
Correspondingly, we can split the integral in Eq. (\ref{newtintegr1}) into three parts:
\begin{equation}
I(r) = I_{0}(r) + I_{\gamma}(r) + I_{R}(r).
\label{intsplit31}
\end{equation}
Here, the integral along the real axis in the $k$-plane
\begin{equation}
 I_0 (r) = I_{0}^{-}(r) + I_{0}^{+}(r),
\label{int0pm1}
\end{equation}
\begin{equation}
I_{0}^{-} (r) =\,\int\limits_{ - \infty }^{ - \varepsilon } {\rm d}k\,k\, \int\limits_0^\infty  {\rm d}r'\,h(r')\,G_l^{(C)(+)}(r,r';k),
\label{int0minus} 
 \end{equation}
\begin{equation}
I_{0}^{+}(r) ={\mathop {\lim }\limits_{\varepsilon  \to 0}}\,\int\limits_\varepsilon ^\infty {\rm d}k\,k\,\int\limits_0^\infty  {\rm d}r'\,h(r')\,G_l^{(C)(+)}(r,r';k). 
\label{int0plus1}
\end{equation}
The integrals along the contours $R$ and $\gamma$ are given by
\begin{equation}
I_R (r) = \int\limits_R {\rm d}k\,k\,\int\limits_0^\infty  {\rm d}r'\,h(r')\,G_l^{(C)( + )}(r,r';k) 
\label{intR1}
\end{equation}
and
\begin{equation}
I_\gamma  (r) = \int\limits_\gamma  {\rm d}k\,k\, \int\limits_0^\infty{\rm d}r'\,h(r')G_l^{(C)( + )}(r,r';k), 
\label{intgamma1}
\end{equation}
correspondingly.
The partial Coulomb Green's function in the configuration space can be written as \cite{newton}
\begin{equation}
G_l^{(C)(+)}(r,r';k) = -\frac{{\varphi _l^{(C)}(k,r_{<})f_{l}^{(C)(+)}(k,r_{>})}}{{L_{l}^{(C)(+)}(k)}}.
\label{Coulgrfunct1}
\end{equation}
Here, $f_{l}^{(C)(\pm )}(k,r)$ are the singular (at $\,r=0$) (Jost) solutions for the $l$-th partial wave of the radial Schr\"odinger equation 
with the pure Coulomb interaction 
\begin{equation}
f_{l}^{(C)(\pm )} (k,r) = e^{\frac{\pi }{2}\eta } W_{ \mp i\eta ,l + 1} ( \mp 2ikr)
\label{singslt1}
\end{equation}
and 
\begin{equation}
L_{l}^{(C)( \pm )} (k) = (2k)^{ - l} e^{\frac{\pi }{2}\eta } e^{ \pm i\frac{\pi }{2}l} \frac{{\Gamma (2l + 2)}}{{\Gamma (l + 1 \pm i\eta )}}
\label{iostfunct1}
\end{equation}
are the Coulomb Jost functions; $\varphi _l^{(C)} (k,r)$ is the regular (at 
$r=0$) solution  given by
\begin{eqnarray}
\varphi _l^{(C)} (k,r) = (2ik)^{- 1}\,  
\left[L_{l}^{(C)(-)}(k)\,f_{l}
^{(C)( + )}(k,r) - L_{l}^{(C)(+)}(k)\,f_{l}^{(C)(-)}(k,r) \right]
\label{regulslt1} \\
=e^{\frac{\pi }{2}\eta } (2k)^{ - l - 1}  
\left[ {\frac{{\Gamma (2l + 2)}}{{\Gamma (l + 1 - i\eta )}}e^{ - i\frac{\pi }{2}(l + 1)} f_{l}^{(C)(+)} (k,r) + \frac{{\Gamma (2l + 2)}}{{\Gamma (l + 1 + i\eta )}}e^{i\frac{\pi }{2}(l + 1)} f_{l}^{(C)(-)} (k,r)} \right]  
\label{regulslt2} \\
=r^{l+1}\,e^{i\,k\,r}\,{}_{1}F_{1}(l=1+i\,\eta,2l+2;-2ikr).
\label{regslhypgm1}
\end{eqnarray}
Also $\eta=Z_{1}\,Z_{2}\,\mu/k$ is the Coulomb parameter of the interacting particles $1$ and $2$, $Z_{i}\,e$ is the charge of particle $i$, $\,\mu$ is the reduced mass of particles $1$ and $2$.
We use the systems of units such that $\hbar=c=1$. 
A physical scattering wave function $\psi_{l}^{(C)}(k,r)$ which is normalized to delta-function is related to the regular solution $\varphi _l^{(C)} (k,r)$ as
\begin{equation}
\psi_{l}^{(C)}(k,r)= e^{i\,\frac{\pi}{2}\,l}\,k\,\frac{\varphi _l^{(C)} (k,r)}{L_{l}^{(C)(+)}(k)}.
\label{physwf1}
\end{equation}

\section{Integral $I(r)$.}
Following Newton's proof we need to show that the integral $I(r)$ over the closed contour $C$, Fig. \ref{fig_1}, is zero. To prove this it is enough 
to show that the Green's function $G_l^{(C)(+)} (r,r';k)$ is a regular function on the integration contour and inside it and that integral $I(r)$ converges.

According to \cite{humblet} the regular Coulomb solution of the Schr\"odinger equation $\varphi _l^{(c)} (k,r)$ is an entire function of $k$
in the complex $k$ plane and is given by a series which absolutely and uniformly converges. 
The singular solution is given by 
\begin{equation}
f_{l( + )}^{(c)} (k,r) = e^{\frac{\pi }{2}\eta } W_{ - i\eta ,l + 1} ( - 2ikr)
\label{jostfunct12}
\end{equation}
where the Whittaker function is determined in a standard way \cite{gradstrizhik}:
\begin{eqnarray}  
W_{ - i\eta ,l + 1} ( - 2ikr) = \frac{{e^{ikr} ( - 2ikr)^{ - i\eta } }}{{\Gamma (l + 1 + i\eta )}}\int\limits_0^\infty  {dte^{ - t} t^{i\eta  + l} \left( {1 + \frac{t}{{( - 2ikr)}}} \right)^{ - i\eta  + l} }, \label{whittaker1}\\ 
\quad |\arg ( - 2ikr)| < \pi, \;\; \label{whittaker2} \\
{\mathop{\rm Im}\nolimits} \,k \ge 0, \,\,\, - i = e^{ - i\pi } e^{i\pi /2}  = e^{ - i\pi /2} 
\label{whittaker3}
\end{eqnarray}
To select the branch of the Whittaker function in the upper half of the $k$ plane ($Im\,k \ge 0$), where the closed contour $C$ lies, 
we impose condition (\ref{whittaker2}) on the argument of the Whittaker function. The Whittaker function is analytical function in $k$ plane with the a branch point singularity at k=0 and the cut going along the negative imaginary semiaxis from $k=0$ to $-i\,\infty$, the blue line in Fig. \ref{fig_1}. Taking into account that $- i = e^{ - i\pi } e^{i\pi /2}  = e^{ - i\pi /2}$, we get that on the bank of the cut in the fourth quadrant $arg\,k= -\pi/2$, and on the left bank of the cut $arg\,k=3/2\,\pi$. 
At such definition $|arg(-2ik)| \ge \pi$. 
Thus the Whittaker function, and, hence, the singular solution are regular functions along and inside the closed integration contour.
We note once again that the integration contour by passes the singularity
of the singular solution at $k=0$. Evidently that the Jost function $L_{l(+)}^{(c)} (k)$ is also regular function along and inside the closed integration contour. The convergence of the integral follows from the behavior of the integrand in the upper half of the $k$ plane. 
It will be shown below when we consider each integral $I_{R}(r),\,I_{0}(r)$ and $I_{\gamma}(r)$. Thus the Newton's integral 
\begin{equation}
I(r)=\oint\limits_{(C)}\,{\rm d}k\,k\,\,\int\limits_0^\infty\,{\rm d}r'\,h(r')\,G_l^{( + )} (r,r';k)=0. \label{intclcont}
\end{equation}
for pure repulsive Coulomb case \footnote[3]{In the presence of the bound states (Coulomb + nuclear potential) the integral $I(r)$ is given by the sum of the residues in the bound state poles for given partial wave $l$.}

\section{Integral $I_{R}(r)$.}

First consider the integral $I_{R}(r)$ over the large 
semicircle $R$; $k \in R: |k| \to \infty$. Following Newton \cite{newton} we split this integral into two parts:
\begin{equation}
I_R (r) = I_{R_{<}}(r) + I_{R_{>}}(r),
\label{intR12}
\end{equation}
where 
\begin{equation}
I_{R_{<}}(r) =- \int\limits_R {\rm d}k\,k\, \int\limits_0^r {\rm d}r'\,h(r')\,
\frac{{\varphi _l^{(C)}(k,r')f_{l}^{(C)(+)}(k,r)}}{{L_{l}^{(C)(+)}(k)}} 
\label{intRs1}, \qquad k \in R,\;\; |k| \to \infty
\end{equation}
and
\begin{equation}
I_{R_{>}}(r) = -\int\limits_R {\rm d}k\,k\, \int\limits_r^\infty  {\rm d}r'\,h(r')\,\frac{{\varphi _l^{(C)}(k,r)f_{l}^{(C)(+)}(k,r')}}{{L_{l}^{(C)(+)} (k)}},\qquad
k \in R, \;\; |k| \to \infty.
\label{intRl1}
\end{equation}
First we consider the integral $I_{R_{<}} (r)$. To evaluate it, we 
replace the regular solution $\varphi _l^{(C)}(k,r)$ by its leading asymptotic term
\begin{eqnarray}
\varphi _l^{(C)} (k,r)  \stackrel{|k| \to \infty}{\approx} (2k)^{ - l - 1} \frac{{\Gamma (2l + 2)}}{{\Gamma (l + 1)}}[e^{ - i\frac{\pi }{2}(l + 1)} e^{i(kr - \eta \ln (2kr))}  \nonumber \\ 
  + e^{i\frac{\pi }{2}(l + 1)} e^{ - i(kr - \eta \ln (2kr))} ].
\label{asympregsol1} 
\end{eqnarray}
Similarly the leading asymptotic term of the singular solution is 
given by
\begin{equation}
f_{l}^{(C)(+)}(k,r) \stackrel{|k| \to \infty}{\approx} e^{i(kr - \eta \ln (2kr))}. 
\label{asympsingsol1}
\end{equation}
Finally the asymptotic behavior of the Jost function is given by
\begin{equation}
L_{l( \pm )}^{(C)} (k) \stackrel{|k| \to \infty}{\approx} (2k)^{ - l} e^{ \pm i\frac{\pi }{2}l} \frac{{\Gamma(2l + 2)}}{{\Gamma (l + 1)}}.
\label{asympiostfunct1}
\end{equation}
Taking into account the expression for the partial-wave Coulomb Green's function, Eq. (\ref{Coulgrfunct1}), we get its leading asymptotic term
\begin{equation}
G_l^{(C)(+)}(r,r';k)= i(2k)^{ - 1} \left[ ( - 1)^l \,e^{ik(r + r')}  - e^{ik(r - r')}  \right].
\label{asympgrfunct1}
\end{equation}
Substituting it into the integral (\ref{intRl1}) gives
\begin{equation}
I_{R_{<}}(r) = i\, \int\limits_R {\rm d}k\,k\,\,\frac{1}{2k}\,\int\limits_0^r\,dr'\,h(r')\, \left[ ( - 1)^l\,e^{ik(r + r')}  - e^{ik(r - r')}  \right].
\label{intrl2}
\end{equation}
We evaluate each term separately using integration by parts. The first 
term containing  $e^{ik(r + r')}$ tends to zero for $\left| k \right| \to \infty$. The second term containing $e^{ik(r' - r)}$
reduces to
\begin{equation}
I_{R_{>}}(r) = \frac{i\,\pi}{2}\,h(r).
\label{intrl3}
\end{equation}
Similarly we get that 
\begin{equation}
I_{R_{<}}(r) =  \frac{i\,\pi}{2}\,h(r).
\label{intrs2}
\end{equation} 
Hence the total integral $I_{R}(r)$ reduces to
\begin{equation}
I_{R}(r)= i\,\pi\,h(r).
\label{intrtot2}
\end{equation}
Thus we arrived exactly at the same result as for the short range interaction \cite{newton}. It is not surprising because the leading asymptotic term of the partial Coulomb Green's function has no trace of the Coulomb interaction and behaves like a leading asymptotic term of the 
free Green's function. 

\section{Integral $I_{0}(r)$}

The integral over the real axis $I_{0}(r)$ is given by the sum (\ref{int0pm1}) of two integrals,
$I_{0}^{(-)}(r)$ and $I_{0}^{(+)}(r)$, Eqs (\ref{int0minus}) and (\ref{int0plus1}).  We split $I_{0}(r)$ into two parts as we did for $I_{R}(r)$:
\begin{equation}
 I_{0_{<}} (r) = I_{0_{<}}^{-}(r) + I_{0_{<}}^{+}(r),
\label{int0less1}
\end{equation}
\begin{equation}
 I_{0_{>}} (r) = I_{0_{>}}^{-}(r) + I_{0_{>}}^{+}(r),
\label{int0large1}
\end{equation}
Here
\begin{equation}
I_{0_{<}}^{\pm}(r) = -{\mathop {\lim }\limits_{\varepsilon  \to 0}}\,\int\limits_{ \pm \infty }^{ \pm \varepsilon } {\rm d}k\,k\, \int\limits_0^r  {dr'h(r')\,\frac{{\varphi _l^{(C)}(k,r')f_{l}^{(C)(+)}(k,r)}}{{L_{l}^{(C)(+)}(k)}}},
\label{int0minus1} 
 \end{equation}
and
\begin{equation}
 I_{0_{>}}^{\pm}(r) = -{\mathop {\lim }\limits_{\varepsilon  \to 0}}\,\int\limits_{\pm \varepsilon}^{\pm \infty}  {\rm d}k\,k\, \int\limits_r^\infty  dr'\,h(r')\,\frac{{\varphi _l^{(C)}(k,r)f_{l}^{(C)(+)}(k,r')}}{{L_{l}^{(C)(+)}(k)}}. 
\label{int0plus11}
\end{equation}
Taking into account that 
$\varphi _l^{(C)} (k,r') = \varphi _l^{(C)} ( - k,r')$ we can rewrite 
the integral $I_{0_{<}}(r)$ as
\begin{equation}
I_{0_{<}}(r) =- \int\limits_0^r {\rm d}r'\,h(r')\left[ {\mathop {\lim }\limits_{\varepsilon  \to 0}} \int\limits_\varepsilon ^\infty {\rm d}k\,k\,\,k\,\varphi _l^{(C)}(k,r')\left\{ \frac{f_{l}^{(C)(+)}(k,r)} {L_{l}^{(C)( + )} (k)} - \frac{f_{l}^{(C)(+)}( - k,r)}{L_{l}^{(C)( + )}( - k)} \right\} \right]. 
\label{inti0small1}
\end{equation}
However 
\begin{equation}
\frac{{f_{l}^{(C)( + )}(-k,r')}}{{L_{l}^{(C)( + )}( - k)}} = \frac{{f_{l}^{(C)(-)}(k,r')}}{{L_{l}^{(C)(-)}(k)}}.
\label{symreliostfunct1}
\end{equation}
Then the integral (\ref{inti0small1}) reduces to
\begin{eqnarray}
I_{0_{<}}(r) = -\int\limits_0^r {\rm d}r'\,h(r')\,\left[ \mathop {\lim }\limits_{\varepsilon  \to 0} \int\limits_\varepsilon ^\infty  {\rm d}k\,k\,\,k\,\varphi_l^{(C)}(k,r')\,\left\{ \frac{f_{l}^{(C)( + )} (k,r)}{L_{l}^{(C)( + )} (k)} - \frac{f_{l}^{(C)(-)}(k,r)}{L_{l}^{(C)(-)}(k)} \right\} \right] 
\label{intioless11} \\
 = -\int\limits_0^r {\rm d}r'\,h(r')\,\left[ \mathop {\lim }\limits_{\varepsilon  \to 0} \int\limits_\varepsilon^\infty  {\rm d}k\,k\,\varphi _{l}^{(C)}(k,r')\, \frac{f_{l}^{(C)( + )}(k,r)\,L_{l}^{(C)( - )}(k) - f_{l}^{(C)( - )}(k,r)\,L_{l}^{(C)( + )}(k)}{L_{l}^{(C)(-)}(k)\,L_{l}^{(C)(+)}(k)}  \right]. 
\label{intioless12}
\end{eqnarray}
Taking into account Eq. (\ref{regulslt1}) we get for $I_{0_{<}}(r)$
\begin{eqnarray}
I_{0_{<}}(r) = -2\,i\,\int\limits_0^r {\rm d}r'\,h(r')\,\left[\mathop {\lim }\limits_{\varepsilon  \to 0} \int\limits_\varepsilon^\infty {\rm d}k\,k^2\,\frac{\varphi_{l}^{(C)} (k,r')\,\varphi_{l}^{(C)}(k,r)}{\left|L_{+}^{(C)} (k) \right|^2} \right]. 
\label{intioless23}
\end{eqnarray}
Similarly we get for $I_{0_{>}}(r)$:
\begin{eqnarray}
I_{0_{>}}(r) = -2\,i\,\int\limits_r^{\infty} {\rm d}r'\,h(r')\,\left[ \mathop {\lim }\limits_{\varepsilon  \to 0} \int\limits_\varepsilon ^\infty  {\rm d}k\,k^2\, \frac{\varphi _{l}^{(C)}(k,r')\,\varphi _{l}^{(C)}(k,r)}{\left|L_{+}^{(C)} (k) \right|^2} \right].  
\label{intiolarge23}
\end{eqnarray}         
Then the integral $I_{0}(r)$ becomes
\begin{equation}
I_{0}(r)= -2i\int\limits_0^{\infty}{\rm d}r'\,h(r')\,\left[\mathop {\lim }\limits_{\varepsilon  \to 0} \int\limits_\varepsilon^\infty {\rm d}k\,k^2\,\frac{\varphi_{l}^{(C)} (k,r')\,\varphi_{l}^{(C)}(k,r)}{\left|L_{+}^{(C)} (k) \right|^2} \right]. 
\label{inti01}
\end{equation}
Let us consider now the limit $\varepsilon \to 0$. As we have mentioned 
the regular solution $\varphi _{l}^{(C)}(k,r)$ is the entire function of $k$ in the finite complex $k$ plane \cite{humblet}. It is given by a series which is absolutely and uniformly converges in the finite $k$ plane \cite{humblet}, i. e. $|\varphi _{l}^{(C)}(k,r)| < A(r)$ in the finite complex $k$ plane including $k=0$.  Now let us consider $1/|L_{+}^{(C)} (k)|^{2}$ for $k \to +0$ \footnote[4]{Limit $k \to +0$ means that $k$ approaches $0$ from the right along the real positive semiaxis.}:
\begin{equation}
1/|L_{+}^{(C)} (k)|^{2}= \frac{1}{[(2l+1)!]^{2}}\,e^{-\pi\,\eta}\,\frac{2\,\pi\,\eta}{e^{\pi\,\eta} - e^{-\pi\,\eta}}\,(2\,k)^{2\,l}\,\prod\limits_{j = 1}^l {(j^2  + \eta ^2 )}\,
\stackrel{k \to +0}{\sim} \, \frac{e^{-2\,\pi\,\eta}}{k} \to 0.
\label{iostfunctm2}
\end{equation}
Thus we can take the limit $\varepsilon \to +0$ in Eq. (\ref{inti01}) 
getting
\begin{equation}
I_{0}(r)= -2i\int\limits_0^{\infty}{\rm d}r'\,h(r')\,\left[ \int\limits_0^\infty {\rm d}k\,k^2\,\frac{\varphi_{l}^{(C)} (k,r')\,\varphi_{l}^{(C)}(k,r)}{\left|L_{+}^{(C)} (k) \right|^2} \right]. 
\label{inti011}
\end{equation}
Again, for the integral along the real axis we got the same results as for the short range interaction. It is not surprisng because eventually we used the symmetry properties of the regular solution under $k \to -k$,
which is the same as for the short range interaction, and limiting behavior of the integrand for $k \to +0$.

\section{Integral $I_{\gamma}(r)$.}

The integral over the small semicircle around $k=0$ is the most difficult part of the proof: the point $k=0$ is a regular point for well behaved short range potentials and singular point for the Coulomb potential. We cannot avoid this integral by just writting $I(r)= I_{R}(r) + \mathop {\lim }\limits_{\varepsilon  \to 0}\,I_{0}(r,\varepsilon)$ and taking the limit $\varepsilon \to 0$ because the integration contour $C$ in this case is not a closed contour and we cannot use Cauchy theorem to evaluate $I(r)$.
We split again the integral $I_{\gamma}(r)$ into two parts:
\begin{equation}
I_{\gamma}(r)= I_{\gamma_{<}}(r) + I_{\gamma_{>}}(r).
\label{intgamma11}
\end{equation}  
Here
\begin{equation}
I_{\gamma_{<}}(r)= -\int\limits_{\gamma } {\rm d}k\,k\, \int\limits_0^r  {\rm d}r'\,h(r')\,\frac{\varphi _l^{(C)}(k,r')\,f_{l}^{(C)(+)}(k,r)}{L_{l}^{(C)(+)}(k)}  
\label{intgammaless1}
\end{equation}
and 
\begin{equation}
I_{\gamma_{>}}(r)= -\int\limits_{\gamma} {\rm d}k\,k\, \int\limits_{r}^\infty  {\rm d}r'\,h(r')\,\frac{\varphi _l^{(C)}(k,r)\,f_{l}^{(C)(+)}(k,r')}{L_{l}^{(C)(+)}(k)}.  
\label{intgammalarge1}
\end{equation}
We will prove that $|I_{\gamma}(r)| \to 0$ as the radius $|k|_{\gamma}$ of the semicircle $\gamma$ goes to zero. We start from consideration of $I_{\gamma_{<}}(r)$.
The regular solution $\varphi _l^{(C)}(k,r)$ does not generate any problems 
for $k \to 0$ for $Im k \ge 0$ because it is entire function of $k$ in the 
finite complex $k$ plane. The singular solution can be written in the form:
\begin{equation}
f_l^{(C)( + )} (k,r) = e^{\pi \eta /2} \frac{ 1}{\Gamma (l + 1 + i\eta )}\,e^{ikr}\,(-2\,i\,k\,r)^{-l}\,\int\limits_0^\infty  {\rm d}t\,e^{ - t}\,t^{2\,l}\,(1 - \frac{{2ikr}}{t})^{ - i\eta  + l}. 
\label{singslt14}
\end{equation}
Letting $z=1-2\,i\,k\,r/t$ we get
\begin{equation}
f_l^{(C)( + )} (k,r) = e^{\pi \eta /2} \frac{ 1}{\Gamma (l + 1 + i\eta )}\,e^{ikr}\,(-2\,i\,k\,r)^{-l}\,\int\limits_0^\infty 
 {\rm d}t\,e^{ - t}\,t^{2\,l}\,z^{ - i\eta  + l}. 
\label{singslt15}
\end{equation}
We can write down 
\begin{equation}
z^{i\,\eta}=z^{-i\frac{\alpha}{k}}= |z|^{-\frac{\alpha}{k}\,\sin \phi}\,
e^{\theta\,\frac{\alpha}{k}\,\cos \phi}\,|z|^{-i\frac{\alpha}{k}\,\cos \phi}\,e^{-i\,\theta\,\frac{\alpha}{k}\,\sin \phi}.
\label{zint1}
\end{equation}
Here, 
\begin{eqnarray}
z= |z|\,e^{i\,\theta}, \qquad \theta = - \arctan \frac{{2k_1 r}}{{t + 2k_2 r}},
\label{z1} \\ 
k=|k|\,e^{i\,\phi}, \qquad \phi=\arctan\,\frac{k_{2}}{k_{1}}= \pi/2 - \arctan\,\frac{k_{1}}{k_{2}}.
\label{k1}
\end{eqnarray}
It is evident that $-\pi/2 \le \theta \le \pi/2\;$ and $\,0 \le \phi \le \pi\;$  because $ -\varepsilon \le k_{1}= Re\,k \le \varepsilon$ and $k_{2} = Im\,k \ge 0\;$ for $\;k \in \gamma$. 
Taking into account that 
\begin{eqnarray}
|z| = \sqrt {(1 + 2\frac{{k_2 r}}{t})^2  + 4\frac{{k_1^2 r^2 }}{{t^2 }}} 
\le  \sqrt {1 + 4\frac{{|k|\, r}}{t} + 4\frac{{|k|^2\,r^2 }}{{t^2 }}} \nonumber\\
\le 1 + \frac{2\,|k|\,r}{t}  \ge 1     \label{modz1}
\end{eqnarray}
and $\frac{\alpha }{{|k|}}\sin \phi \ge 0$ for $k_{2} \ge 0$ we get  $| z |^{ - \frac{\alpha }{{|k|}}\sin \phi } \le 1$.
Besides 
\begin{eqnarray}
\theta\,\frac{\alpha}{k}\,\cos \phi=-\frac{\alpha}{k}\,\arctan(\frac{{2k_1 r}}{{t + 2k_2 r}})\,\sin(\arctan\frac{k_{1}}{k_{2}}) \nonumber\\
 = -\frac{\alpha}{k}\,\arctan(\frac{{2k_1 r}}{{t + 2k_2 r}})\,\frac{k_{1}}{\sqrt{k_{1}^{2}+ k_{2}^{2}}} \le 0,  \qquad k \in \gamma.
\end{eqnarray}
Then
\begin{equation}
\left| {z^{ - i\,\eta} } \right| \le \left| {\left| z \right|^{ - \frac{\alpha }{{|k|}}\sin \varphi } e^{\theta \frac{\alpha }{{|k|}}\cos \varphi } } \right| \le 1.
\label{ineq1}
\end{equation}
Correspondingly for $f_l^{(C)( + )} (k,r)$ we get
\begin{eqnarray}
|f_l^{(C)( + )} (k,r)| \le |e^{\pi \eta /2} \frac{ 1}{\Gamma (l + 1 + i\eta )}\,(-2\,i\,k\,r)^{-l}|\,\int\limits_0^\infty 
 {\rm d}t\,e^{ - t}\,t^{2\,l} |z^{l}|   \nonumber\\
\le  |e^{\pi \eta /2}\,\frac{ 1}{\Gamma (l + 1 + i\eta )}\,(-2\,i\,k\,r)^{-l}|\,\int\limits_0^\infty 
 {\rm d}t\,e^{ - t}\,t^{l}\,(t+ 2\,|k|\,r)^{l}. 
\label{singslt25}
\end{eqnarray}
Note that in the integrand we majorized $|z|^{l}$ by $[(t+2\,|k|\,r)/t]^{l}$following Eq. (\ref{modz1}). We also took into account that for $k \in \gamma$ ($k_{2} \ge 0$) $\;e^{i\,k\,r}| \le 1$. It is evident that the integral over $t$ converges and can be uniformly majorized in the finite $k$ plane. Let 
\begin{equation}
J(|k|\,r)=\int\limits_0^\infty {\rm d}t\,e^{ - t}\,t^{l}\,(t+ 2\,|k|\,r)^{l}.
\label{intj1}
\end{equation}
Let $s$ be a semicircle in the upper half $k$ plane with the radius $|k|_{s} > |k|_{\gamma}$. Then for all $k \in s \;J(|k|\,r) \le J(|k|_{s}\,r) $ and $|f_l^{(C)( + )} (k,r)|$ can be majorized as follows:
\begin{equation}
|f_l^{(C)( + )} (k,r)| \stackrel{|k| \to 0}{\le}  |e^{\pi \eta /2}\,\frac{ 1}{\Gamma (l + 1 + i\eta )}\,(-2\,i\,k\,r)^{-l}|\,J(|k|_{s}\,r). 
\label{singslt251}
\end{equation}
Taking into account Eq. (\ref{iostfunct1}) we get 
\begin{equation}
|\frac{|f_l^{(C)( + )}(k,r)|}{L_{l}^{(C)( \pm )}(k)}| \stackrel{|k| \to 0}
{\le} \frac{1}{\Gamma(2L+2)}\,J(|k|_{s}\,r).
\label{singsoliostfnct1}
\end{equation}
Then it is straigtforward to see from Eq. (\ref{intgammaless1}) that
\begin{equation}
|I_{\gamma_{<}}(r)| = |\int\limits_{\gamma } {\rm d}k\,k\, \int\limits_0^r  {\rm d}r'\,h(r')\,\frac{\varphi _l^{(C)}(k,r')\,f_{l}^{(C)(+)}(k,r)}{L_{l}^{(C)(+)}(k)}| \le 2\,\frac{1}{\Gamma(2L+2)}\,J(|k|_{s}\,r)\,|k|\,|\int\limits_{0}^{r} {\rm d}r'\,h(r')\,A(r')| \to 0,
\label{intgammaless12}
\end{equation}
where $A(r) \ge |\varepsilon_{l}^{(C)}(k,r)|$ is the majorizing function for $k \in s$. We also took into account that $\gamma$ is $|\int\limits_{\gamma} {\rm d}k\,k\,|=2\,|k|$ where $\gamma$ is the semircle.
Similarly we prove that 
\begin{equation}
|I_{\gamma_{>}}(r)| = |\int\limits_{\gamma} {\rm d}k\,k\, \int\limits_{r}^\infty  {\rm d}r'\,h(r')\,\frac{\varphi _l^{(C)}(k,r)\,f_{l}^{(C)(+)}(k,r')}{L_{l}^{(C)(+)}(k)}| \le  2\,\frac{1}{\Gamma(2L+2)}\,\,A(r)\,|k|\,|\int\limits_{r}^{\infty} {\rm d}r'\,h(r')\,J(|k|_{s}\,r')| \to 0.  
\label{intgammalarge12}
\end{equation}
Note that assumption that $h(r) \in L^{2}$ is enough to provide the convergence of the radial integral in Eq. (\ref{intgammalarge12}).

Thus we proved that $|I_{\gamma}(r)| \le ||I_{\gamma_{<}}(r)| + |I_{\gamma_{>}}(r)| \stackrel{|k| \to 0}{\to} 0$. 

\section{Completeness of the Coulomb scattering wave functions.}
Now we can return to Eq. (\ref{intsplit31}). Replacing $I(r)$ and $I_{\gamma}(r)$ by zero, $I_{R}(r)$ by Eq. (\ref{intrtot2}) and $I_{0}(r)$ by Eq. (\ref{inti011}) we get
\begin{equation}
h(r)= \frac{2}{\pi}\,\int\limits_0^{\infty}{\rm d}r'\,h(r')\,\left[ \int\limits_0^\infty {\rm d}k\,k^2\,\frac{\varphi_{l}^{(C)} (k,r')\,\varphi_{l}^{(C)}(k,r)}{\left|L_{+}^{(C)}(k) \right|^2} \right].
\label{finequat1}
\end{equation}
From this equation we may conclude that the Coulomb scattering wave functions for repulsive Coulomb interaction satisfy the completeness relationship:
\begin{equation}
\delta(r'-r)= \frac{2}{\pi}\, \int\limits_0^\infty {\rm d}k\,k^2\,\frac{\varphi_{l}^{(C)} (k,r')\,\varphi_{l}^{(C)}(k,r)}{\left|L_{+}^{(C)}(k) \right|^2}.
\label{deltafunct1}
\end{equation}
Introducing the spectral function $\rho(E)$ we can rewrite 
the completeness relationship (\ref{deltafunct1}) in the form:
\begin{equation}
\delta(r'-r)=  \int\limits_0^\infty {\rm d}\rho(E)\,\varphi_{l}^{(C)} (k,r')\,\varphi_{l}^{(C)}(k,r),
\label{deltafunct2}
\end{equation} 
where 
\begin{equation}
\frac{{\rm d}\rho}{{\rm d}E}= \cases{\frac{2\,\mu\,k}{\pi}\, \left|L_{+}^{(C)}(k) \right|^{-2},  \qquad  E \ge 0, \cr
0, \qquad\qquad\qquad\qquad\quad  E <0.  \cr}
\label{spectrfunct1}
\end{equation}
Using the physical scattering wave function (\ref{physwf1}) we can rewrite 
completeness relationship (\ref{deltafunct1}) in the standard form
\begin{equation}
\delta(r'-r)= \frac{2}{\pi}\, \int\limits_0^\infty {\rm d}k\,\psi_{l}^{(C)*} (k,r')\,\psi_{l}^{(C)}(k,r).
\label{deltafunctphyswf1}
\end{equation}
Thus for the repulsive Coulomb interaction we got the same 
result as Newton \cite{newton60,newton} for the short range interactions:
the system of the physical scattering wave functions ${\it S}= \{\psi_{l}^{(C)}(k,r)\}$ forms a complete set. 

\section{Berggren's method and normalization of the resonance Gamow functions
for charged particles}

The completeness of the eigenfunctions of the two-body Hamiltonian with the short range interaction and the formal proof of the completeness of the Coulomb scattering wave functions presented here allows us to claim that the eigenfunctions of the Hamiltonian for the repulsive Coulomb + short range (nuclear) potential form a complete set \footnote[5]{We exclude potentials supporting the bound state at $E=0$.}:
\begin{equation}
\delta(r'-r)= \sum\limits_n\,\varphi _{nl} (r')\, \varphi _{nl} (r)\,
+\frac{2}{\pi}\, \int\limits_0^\infty {\rm d}k\,k^2\,\frac{\varphi_{l}^{(C)} (k,r')\,\varphi_{l}^{(C)}(k,r)}{\left|L_{+}^{(C)}(k) \right|^2}.
\label{deltafunct12}
\end{equation}
Here $\varepsilon_{nl}$ is the normalized to unity bound state wave function 
with the principal quantum numer $n$ in the partial wave $l$. In  Berggren's method \cite{berggren68} the complete set
of the bound states and continuum states for real positive energies $E \ge 0$ can be redefined by including the resonant states. A new complete system consists of the discrete states, bound and resonant, and continuum states,
scattering states for real positive and complex energies. To normalize the resonant states Berggren intdoduced the dual basis and used Zel'dovich's regularization procedure \cite{zeldovich60}.  However, Zel'dovich's regularization procedure has been shown to work only for the resonances
for the short range potentials. We will show here that Zel'dovich's regularization works also for the resonant Gamow states with the Coulomb tail.   

Let $\varphi_{nl}^{R}(r)$ stands for the Gamow wave function describing the resonant state in a system of two charged particles interacting via the sum of the repulsive Coulomb + nuclear potential. The dual Gamow function
${\tilde \varphi}_{nl}^{R}(r)$ satisfies the condition: 
${\tilde \varphi}_{nl}^{R*}(r)= \varphi_{nl}^{R}(r)$. 
Asymptotic behavior of the Gamow function is given by
\begin{equation}
\varphi_{nl}^{R}(r) \stackrel{r \to \infty}{\approx}
b_{nl}\,\frac{e^{i\,k^{R}\,r}}{r^{i\,\eta^{R}}},
\label{gamfunctasympt1}
\end{equation}
where $b_{nl}$ is the amplitude of the tail of the resonant Gamow function, $k^{R}= k_{1} + i\,k_{2}$ is the momentum of the resonance, $k_{1}=Rek >0$
and $k_{2}= Im k <0$; $\,\eta^{R}= Z_{1}\,Z_{2}\,e^{2}\,\mu/k^{R}$ is the Coulomb parameter for the resonant state.
The radial factor $r^{i\,\eta^{R}}$ in the denominator appears due to the Coulomb barrier.
Zeldovich's normalization condition of the Gamow function is given by 
\begin{equation}
N=\mathop {\lim }\limits_{\beta \to 0}\, \int\limits_{0}^{\infty} {\rm d}r\,\varphi_{nl}^{R}(r)\,{\tilde \varphi}_{nl}^{R*}(r)\,e^{-\beta\,r^{2}}.
\label{gamowstnrm1}
\end{equation}
The introduction of the regularization factor $\exp(-\beta\,r^{2})$ is neccessary to provide the convergence of the normalization integral
for $ r \to \infty$ which otherwise diverges because $\exp(i\,k^{R}\,r)=
\exp(i\,k_{1}\,r - k_{2}\,r)$.  
We split the integral (\ref{gamowstnrm1}) into two parts:
\begin{eqnarray}
N=\mathop {\lim }\limits_{\beta \to 0}\,[ \int\limits_{0}^{A} {\rm d}r\,\varphi_{nl}^{R}(r)\,{\tilde \varphi}_{nl}^{R*}(r)\,e^{-\beta\,r^{2}} +
b_{nl}\int\limits_{A}^{\infty} {\rm d}r\,\frac{e^{i\,2\,k^{R}\,r}}{r^{i\,2\,\eta^{R}}}\,e^{-\beta\,r^{2}}]  \nonumber\\
= \int\limits_{0}^{A} {\rm d}r\,\varphi_{nl}^{R}(r)\,{\tilde \varphi}_{nl}^{R*}(r)+
b_{nl}\,\mathop {\lim }\limits_{\beta \to 0}\, \int\limits_{A}^{\infty} {\rm d}r\,\frac{e^{i\,2\,k^{R}\,r}}{r^{i\,2\,\eta^{R}}}\,e^{-\beta\,r^{2}}.
\label{gamowstnrm2}
\end{eqnarray}
$A$ is assumed to be large enough to approximate at $r > A$ the Gamow function by its leading asymptotic term (\ref{gamfunctasympt1}).
Evidently, the integral over a finite interval $r \le A$ converges and we can take limit $\beta \to 0$ in this integral. We need to prove that the second integral can also be determined in the limit $\beta \to 0$.  
Taking into account that
\begin{equation}
\eta^{R}= \frac{\alpha}{k^{R}}= \frac{\alpha}{(k_{1})^{2}+(k_{2})^{2}}(k_{1}- i\,k_{2})= \lambda - i\,\delta, 
\label{etarepr1}
\end{equation}
where $\alpha= Z_{1}\,Z_{2}\,e^{2}\,\mu$, and $\lambda > 0 $ and $\delta < 0$.
Hence 
\begin{equation}
N_{ext}= b_{nl}\,\mathop {\lim }\limits_{\beta \to 0}\, \int\limits_{A}^{\infty} {\rm d}r\,e^{i\,2\,k_{1}\,r}\,e^{-2\,k_{2}\,r}\,r^{-2\,i\,\lambda}\,r^{-2\,\delta}  \,e^{-\beta\,r^{2}}.
\label{normext1}
\end{equation}
For neutral particles there is only one exponentially diverging factor $\exp(-2\,k_{2}\,r)$ (for $ r \to \infty$)  in the integrand.
Zel'dovich's regularization method was applied to regularize  such integrals.
This procedure works for $k_{1} > k_{2}$.
The presence of the oscillating exponential $\exp(i\,2\,k_{1}\,r)$ is crucial for Zel'dovich's regularization. 
However for charged particles due to the Coulomb barrier at complex momentum we have an additional diverging factor in the integrand, $r^{-2\,\delta}$. We will show now that Zel'dovich method works for the charged particles also. We rewrite Eq. (\ref{normext1}) as the sum of two terms:
\begin{equation}
N_{ext}= N_{1}+N_{2}=-b_{nl}\,\mathop {\lim }\limits_{\beta \to 0}\, \int\limits_{0}^{A} {\rm d}r\,e^{i\,2\,k^{R}\,r}\,r^{-2\,i\,\eta^{R}}\,e^{-\beta\,r^{2}} + b_{nl}\,\mathop {\lim }\limits_{\beta \to 0}\, \int\limits_{0}^{\infty} {\rm d}r\,e^{i\,2\,k^{R}\,r}\,r^{-2\,i\,\eta^{R}}\,e^{-\beta\,r^{2}}.
\label{normext12}
\end{equation}
The first term in this equation converges and we can take limit $\beta \to 0$ in this term. It is enough to consider the second term only. 
The second integral can be taken analytically using Eqs. (3.462(1)) and (9.246) \cite{gradstrizhik}:
\begin{eqnarray}
\mathop {\lim }\limits_{\beta \to 0}\, \int\limits_{0}^{\infty} {\rm d}r\,e^{i\,2\,k\,r}\,r^{-2\,i\,\eta}\,e^{-\beta\,r^{2}} 
&=& \mathop {\lim }\limits_{\beta \to 0}\,(2\beta )^{\frac{{2\,i\,\eta  - 1}}{2}}\,\Gamma ( - 2\,i\,\eta + 1)\,e^{ - \frac{k^2}{2\beta}}\, D_{2\,i\,\eta  - 1} (-\frac{2\,i\,k}{\sqrt {2\beta}}) \nonumber\\
&&=\Gamma ( - 2\,i\,\eta + 1)( - i\,2\,k)^{2\,i\,\eta  - 1}=J(k).
\label{jregint1}
\end{eqnarray}
Here, $D_{\nu}(x)$ is the parabolic cylinder function. For $Im k <0$ and $Re(-2\,i\,\eta) > 0$ the integral on the left hand-side of this equation determined 
only for $\beta >0$. Meantime function $J(k)$ is an analytical function of $k$ in the whole complex $k$ plane except for the singular points at $2\,i\,\eta =n$, where $n=1,2...$, and $k=0$. Hence function $J(k)$ can be considered as an analytical continuation of the integral 
$ \int\limits_{0}^{\infty} {\rm d}r\,e^{i\,2\,k\,r}\,r^{-2\,i\,\eta}$
into region $Im k <0$. 
Thus we have shown that the Zel'dovich's regularization can be used to determine the norm of the Gamow functions for charged particles. 

\section{Summary} 

We presented a formal proof of the completeness of the eigenfunctions of the 
two-body Hamiltonian with repulsive Coulomb potential using the Newton's integral \cite{newton60,newton}. The most delicate point was to
investigate the behavior of the integrand near the singular point $k=0$ and to prove that the contribution from a small semicircle $\gamma$ around $k=0$ goes to zero as its radius $|k|_{\gamma} \to 0$. The presented proof allows us to claim that a system of the eigenfunctions ${\it S}=\{\varphi_{nl}(r),\,\psi_{l}(k,r)\},\;0 <n < N,\, 0 < k < \infty$, of the  of the self-adjoined Hamiltonian with the potential given by the sum of the Coulomb and nuclear interactions is also complete.

Recently the Berggren's method \cite{berggren68} has been used in the so-called Gamow shell model \cite{michel}. The Berggren's method can also be used for the Green's function spectral decomposition in nuclear reaction theory with charged particles \cite{berggren68}. Our proof validates  application of the Berggren's method \cite{berggren68} for charged particles. We have also demonstrated that the Gamow resonant states with the Coulomb tail can be normalized using Zel'dovich's regularization method \cite{zeldovich60}. It is a crucial point for application of the Berggren's technique. Note that Zel'dovich's regularization method is not unique and other regularization techniques can be used.

\section{Acknowledgments}
This work was supported by the U.\,S. DOE under Grant No.\@ DE-FG02-93ER40773 and the U.\,S. NSF under Grant No. \@ PHY-0140343.
M. A. expresses his thanks to National Science Foundation for support through the REU Summer program 2005 and the Cyclotron Institute for hospitality during his stay in Summer 2005.


\begin{thebibliography}{99}
\bibitem{newton60} R. G. Newton, J. Math Phys.,  \textbf{1}, 319 (1960).
\bibitem{newton} R. G. Newton, Scattering Theory of Waves and Particles, 2nd ed., Springer-Verlag, Heidelberg, 1982.
\bibitem{ment65} Yu. L. Mentkovsky, Nucl. Phys. {\bf 65}, 673 (1965).
\bibitem{humblet} J. Humblet, Ann. Phys. {\bf 155}, 461 (1984).
\bibitem{gradstrizhik} I. S. Gradshteyn and I. M. Ryzhik, Tables of Integrals, Series and Products, 4-th edition, Ed. A. Jeffrey, Academic Press, New York-London-Toronto-Sydney-San-Francisco, 1980.
\bibitem{berggren68} T. Berggren, Nucl. Phys. {\bf A109}, 265 (1968).
\bibitem{zeldovich60} Ya. B. Zel'dovich, ZhETF {\bf 39}, 776 (1960).
\bibitem{michel} N. Michel, W. Nazarewicz, M. Ploszajczak and K. Bennaceur, Phys. Rev. Lett. {\bf 89}, 042502 (2002).
\end{thebibliography}
\end{document}